\documentstyle[12pt]{article}
\begin{document}
\begin{center}
\begin{large}
\mbox{}\vspace{5mm}

{\bf Stability of a hot two-temperature accretion disc with advection}\\
[5mm]
\end{large}
\end{center}
\begin{center}
Xue-Bing Wu \\
        Institute of Theoretical Physics, Chinese Academy of Sciences, Beijing 
        100080, China \& also Beijing Astronomical Observatory, Chinese 
        Academy of Sciences, Beijing 100080, China 
\\ 
 email: wuxb@itp.ac.cn\\[8mm]
\end{center}
\noindent
Contact addresses:\\
\begin{center}
(Except from April 20, 1997 to August 4, 1997)\\
Dr. Xue-Bing Wu\\
Institute of Theoretical Physics\\
Chinese Academy of Sciences\\
Beijing 100080\\
China\\[3mm]
email: wuxb@itp.ac.cn\\
Fax: +86-10-6256-2587\\
Telephone: +86-10-6257-8296\\[1cm]
(Within from April 20, 1997 to August 4, 1997)\\
Dr. Xue-Bing Wu\\
Max-Planck-Institut f\"ur Extraterrestrische Physik\\
85740 Garching Bei M\"unchen\\
Germany\\[3mm]
email: xwu@mpe-garching.mpg.de\\
Fax: +49-89-3299-3569\\Telephone: +49-89-3299-3838
\end{center}
\newpage
\begin{center}
{\bf Abstract}
\end{center}
The effects of radial advection and thermal diffusion were considered in 
investigating the linear stability of an optically thin, two-temperature 
accretion 
disc. 
If the disc has only very little advection, we proved that the thermal 
instability exists when the disc is geometrically thin. But it dispears in 
a geometrically slim disc if the thermal diffusion 
was
considered. Moreover, if the disc is advection dominated, the thermal 
instability does not exist. In addition, we found that the instabilities of 
inertial-acoustic modes exist only in a geometrically thin disc or an 
advection-dominated disc with low Mach number, whereas the Lightman \& Eardley
viscous instability always dispears in a two-temperature disc. A simple 
comparison also showed that an optically thin, bremsstrahlung cooling dominated 
disc is generally more thermally unstable than a two-temperature disc if it is
not advection-dominated.\\[5mm]

\newpage
\section{Introduction}

The optically thin, two-temperature accretion disc model was suggested in 1970s
to explain the hard X-ray spectra observed in black hole candidates such as
Cyg X-1 (Shapiro, Lightman \& Eardley 1976, hereafter SLE). The standard 
geometrically thin, optically thick accretion
 disc model, proposed by Shakura \& Sunyaev (1973), was unable to account 
 for 
them. In a two-temperature disc, the temperature of electron is about $10^9$K 
and the ion temperature is one or two orders higher. Therefore, the 
two-temperature disc has much higher temperature than the cool, optically 
thick disc and can produce the observed hard X-ray spectrum above $\sim$8 
keV.  In last two decades, SLE model has been widely studied and applied in 
modeling of X-ray binaries and active galactic nuclei(e.g., Kusunose \& 
Takahara 1988, 1989; White \& Lightman 1989; Wandel \& Liang 1991; Luo \& 
Liang 1994). However, the early work of Pringle (1976) and Piran (1978) 
indicated that the two-temperature disc model is still thermally unstable. 
As in the standard thin disc model, which is both thermally and viscously 
unstable in the inner region (Shakura \& Sunyaev 1976; Lightman \& Eardley 1974), the rapid 
growing of instability may result in the break
down of the disc equilibrium and make it unlikely to be
the real configuration of accretion flow. 

In most previous studies we mentioned above, the disc model was relatively 
simple and many effects were ignored.  Recently, the effects of radial advection were 
extensively studied in both optically thick and optically thin accretion 
discs (Abramowicz et al. 1988; Kato, Honma \& Matsumoto
1988; Narayan \& Yi 1994, 1995a,b; Abramowicz et al. 
1995; Chen et al. 1995; Chen 1995; Misra \& Melia 1996; Nakamura et al. 1996). 
If the radiative cooling 
is not efficient 
to balance the viscous heating in
the disc, some energy will be advected inward and the advection will be
not negligible. It has been suggested 
both the disc structure and the stability properties of an accretion disc 
with advection term included
are different from those in previous models where advection was totally neglected.
By analyzing the $\dot{M}(\Sigma)$ slope of disc structure and
comparing the cooling and heating rates near each equilibrium curve, Chen et al. 
(1995), Abramowicz et al. (1995) and Narayan \& Yi (1995b) have suggested that an advection-dominated disc is both thermally and 
viscously stable whether the disc is optically 
thin or optically thick. However, such a stability analysis, as well as Piran's
criteria, 
can only be applied
to the case with long-wavelength perturbations (Chen 1996). More recently, Kato, 
Abramowicz \& Chen (1996) performed an analytic stability analysis of the 
advection-dominated discs by considering the short wavelength perturbations.
They 
indicated that the advection-dominated disc is thermally stable if it is 
optically thin but thermally unstable if it is optically thick. This result 
was confirmed by a subsequent study of Wu \& Li (1996), who considered not 
only the stability of thermal mode, but also the stability of viscous mode 
and inertial-acoustic modes. They also concluded that the thermal diffusion 
has a significant contribution to stabilize an optically thin, 
advection-dominated disc but
enhance the thermal instability of an optically thick, advection-dominated disc.

Although some authors have mentioned that an optically thin, two-temperature 
disc may be thermally stable if it is advection-dominated (e.g.,
Narayan \& Yi 1995b), however, no detailed
stability analysis has been done so far to consider the effects of advection on 
the stability of a two-temperature accretion disc. For example, in the 
recent work of Wu \& Li (1996), only the optically thin disc with 
bremsstrahlung radiative cooling was addressed. We noted that advection may 
be quite important in a two-temperature disc as calculated by some authors 
(for example, see Esin et al. 1996). Moreover, a hot two-temperature
disc is probably not geometrically thin and the Piran's criteria may not be
applicable in this case. Therefore, in this paper we performed a 
detailed linear study to investigate the stability of a hot two-temperature 
disc, considering 
both the
effects of advection and thermal diffusion. Besides the stability of thermal 
mode, the stability of viscous mode and that of inertial-acoustic modes have also 
been investigated.

\section[]{Basic equations}

The simplified equations which describe the equilibrium structure of a 
two-temperature disc has been given by SLE. However, in order to study the 
stability properties, we must involve the detailed time-dependent hydrodynamic 
equations. In addition, because we will investigate the influences of some 
terms such as the radial viscous force, advection and thermal diffusion on 
the disc stability, we should also included them in the equations.  
Ignoring the self-gravitation and adopting the pseudo-Newtonian
potential (Paczy\'nski \& Wiita 1980), $\Psi=-GM/(R-r_{g})$, we can write
the vertical 
integrated time-dependent equations in a cylindrical system of coordinates
 as follows:
\begin{equation} 
\frac{\partial{\Sigma}}{\partial{t}}+\frac{1}{r}\frac{\partial}{\partial{r}}
(r\Sigma V_{r})=0, 
\end{equation}
\begin{equation}
\Sigma\frac{\partial{V_{r}}}{\partial{t}}+\Sigma V_{r}\frac{\partial{V_{r}}}
{\partial{r}}-\Sigma(\Omega^{2}-{\Omega_{k}}^{2})r=-2\frac{\partial{(Hp)}}{\partial{r}}
+F_{\nu},
\end{equation}
\begin{equation}
\Sigma r^{3}\frac{\partial{\Omega}}{\partial{t}}+\Sigma rV_{r}\frac{\partial{
(r^{2}\Omega)}}
{\partial{r}}=\frac{\partial}{\partial{r}}(\Sigma \nu r^{3} \frac{\partial{\Omega}}
{\partial{r}}),
\end{equation}
\begin{equation}
C_{v}[\Sigma\frac{\partial{T}}{\partial{t}}+\Sigma V_{r}\frac{\partial{T}}{
\partial{r}}-(\Gamma_{3}-1)T(\frac{\partial{\Sigma}}{\partial{t}}+V_{r}\frac
{\partial{\Sigma}}{\partial{r}})]=\Sigma \nu (r\frac{\partial{\Omega}}{\partial
{r}})^{2}-Q_{-}+Q_{t},
\end{equation}
where $ V_{r}$, $\Omega$ are the radial velocity and angular velocity, $\Omega_{k}$ is the Keplerian angular 
velocity, given by ${\Omega_{k}}^{2}=(\frac{\partial{\Psi}}{r \partial{r}})_{z=0}
$. $p$, $T$, 
$\Sigma$  and H are the total pressure, temperature, surface density and the disc height, $C_{v}$ and
$\Gamma_{3}$ are the heat capacity per unit mass and a quantity associated with 
$\beta$, the ratio of gas to total pressure. For an optically thin, two-temperature disc, $\beta=1$, $C_{v}=3p/(2\rho T)$,
$\Gamma_{3}=5/3$ and $p=\rho k(T_i + T_e)/m_p$, where $T_i$, $T_e$ and $m_p$ are the ion, electron temperatures and proton mass. $F_{\nu}$ is the radial viscous force, which is often neglected in geometrically 
thin accretion discs but is perhaps not negligible and should be
considered 
in accretion discs with
advection (Narayan \& Yi 1995a). It is given by (Papaloizou \& Stanley 1986)
\begin{equation}
F_{\nu}=\frac{\partial}{\partial{r}}[\frac{1}{2} \frac{\nu_{r}\Sigma}{r} \frac
{\partial{(rV_{r})}}{\partial{r}}]-\frac{2V_{r}}{r}\frac{\partial{(\nu_{r}\Sigma)}}
{\partial{r}},
\end{equation}
where $\nu_{r}$ is the kinematic viscosity acting in the radial direction. In this
paper we take $\nu_{r}=\nu$, where $\nu$ is the viscosity acting in the azimuthal
direction and is expressed as the standard $\alpha$ prescription (Shakura \&
Sunyaev 1973), $\nu=\alpha c_{s}H$. $c_{s}$ is the local sound speed defined by
$c_{s}=p/\rho$. The disc height is given by
$H=c_{s}/\Omega_{k}$. $Q_{-}$ at the right side of Eq. (4) represents the radiative 
cooling. For an optically thin, one-temperature disc, the radiative cooling mechanism 
is usually taken as thermal bremsstrahlung. However, for a
two-temperature disc, the cooling mechanism is unsaturated Comptonization through
the loss of energy of electrons. In this case, the cooling rate can be expressed as 
(see also SLE)
\begin{equation}
Q_- =(4kT_e /m_e c^2)\rho H \kappa_{es} U_r c
\end{equation}
where $m_e$ is the electron mass, $\kappa_{es}=0.4cm^2g^{-1}$ is the electron scattering opacity. $U_r$ is the radiation energy density of soft photons, which
we assume, for simplicity, does not change on timescale short compared to $\Omega^{-1}$ (Pringle 1976).
Because the ions and electrons are coupled by collisional energy exchange, the 
loss of energy of electrons can be balanced by the energy capture from ions. The exchange rate is
\begin{equation}
Q_{i-}=(3/2)\rho H \nu_E k(T_i-T_e)/m_p
\end{equation}
where $\nu_E$ is the electron-ion coupling rate, and can be approximated by 
$\nu_E=2.4\times 10^{21}ln \Lambda {\rho}{T_e}^{-3/2}$ (Spitzer 1962) where the Coulomb 
logarithm $ln \Lambda$ is about 15 (SLE). Taking $Q_-=Q_{i-}$ and $T_i >> T_e$, we can 
get $Q_-\propto \Sigma^{7/5}{T_i}^{1/5}$.
$Q_{t}$ at the right side of Eq. (4)
represents the thermal diffusion defined as $Q_{t}=\nabla\cdot(K
\nabla T)$, where $K$ is the vertical integrated thermal conductivity
given by $K=\Sigma C_{v}\nu =\alpha f\Omega_{k}H^{3}p/T$, where $f=3(8-7\beta)
f_{*}$ and $f_{*}$ is of the order of unity (Kato et al.
1996; Wu \& Li 1996). In this paper, we take $f=3$ since a two-temperature disc is usually gas pressure dominated and $\beta \simeq 1$.

 The equilibria of a two-temperature disc can be calculated by numerically solving
 above equations with $\partial/\partial t$=0. Ignoring the influences of
 radial viscous force and thermal diffusion, the disc equilibria have been
 calculated by many authors (e.g., Narayan \& Yi 1995b;
Chen et al. 1995; Nakamura et al. 1996). In this paper, we assume the 
influences of radial viscous force and thermal diffusion on the disc equilibria
 are small, and refer to the solution of the disc structure obtained
 by Narayan \& Yi (1995b). We also restrict our stability analyses within the validity of local approximation and vertical integrated equations, which requires that $\lambda<r$ and $kV_r<\Omega_k$. These two inequalities can be summerized as $\frac{r}{H}>\frac{\lambda}{H}>2\pi\alpha\frac{H}{r}$, which is well satisfied for a geometrically thin
accretion disc in a wide range of perturbation wavelength but for a geometrically slim disc only when the  perturbation wavelength is short and viscosity efficient $\alpha$ is sufficiently small (Kato et al., 1996; Wu \& Li 1996). The vertical hydrostatic balance can be realized if the time scale associated with the perturbations is longer than the dynamical time scale. Although for a geometrically thick disc, the perturbation wavelength can be smaller than the disc height ($\lambda<H$), which means that the perturbations are also local in the vertical direction (Kato et al. 1997), the consideration of the local vertical perturbations for
a two temperature disc does not change the stability properties very much (Yamasaki 1997).

\section{Linear perturbations and dispersion relation}
In this section, we consider the linear perturbations to the two-temperature disc. The radial perturbations of $V_{r}, \Omega, \Sigma$ and $T$ are assumed of the
 form $(
\delta V_{r}, \delta \Omega, \delta \Sigma, \delta T)\sim e^{i(\omega t-kr)}$,
where $k$ is the perturbation wavenumber defined by $k=2\pi/\lambda$, $\lambda$ is
the perturbation wavelength. Taking the perturbed quantities to the basic equations given 
in last section and considering the local approximation, we can obtain following perturbed 
equations:
\begin{equation}
\tilde{\sigma}\frac{\delta\Sigma}{\Sigma}-i \frac{\epsilon}{\tilde{H}} \frac{\delta 
V_{r}}{\Omega_{k}r}=0,
\end{equation}
\begin{equation}
-i \epsilon \tilde{H} \frac{\delta \Sigma}{\Sigma}+(\tilde{\sigma}+\frac{4}{3}
\alpha \epsilon^{2})\frac{\delta V_{r}}{\Omega_{k} r}-2\tilde{\Omega}\frac{\delta
 \Omega}{\Omega_{k}}-i \epsilon \tilde{H} \frac{\delta T}{T}=0,
\end{equation}
\begin{equation}
 i \alpha \epsilon\tilde{H}g\frac{\delta \Sigma}{\Sigma}+\frac{{\tilde{\chi}}^{2}}
 {2\tilde{\Omega}}\frac{\delta V_{r}}{\Omega_{k} r}+(\tilde{\sigma}+\alpha\epsilon^{2}
 )\frac{\delta \Omega}{\Omega_{k}}+i\alpha\epsilon\tilde{H}g\frac{\delta T}
 {T}=0,
\end{equation}
\begin{equation}
 -(\tilde{\sigma}+\alpha (-\frac{2}{5}+\frac{7}{5}q) g^{2})\frac{\delta\Sigma}{\Sigma}-\frac{\alpha q
 g^{2}}{m\tilde{H}}\frac{\delta V_{r}}{\Omega_{k} r}+\frac{2i\alpha\epsilon g}
 {\tilde{H}}\frac{\delta{\Omega}}{\Omega_{k}}+(\frac{3}{2}\tilde{\sigma}+\alpha (\frac{4}{5}+\frac{1}{5}q)
 g^{2}+\alpha f \epsilon^{2})\frac{\delta T}{T}=0,
\end{equation}
where $\tilde{\sigma}=\sigma/\Omega_{k}$, and $\sigma=i(\omega-kV_{r})$. 
$\tilde 
{\Omega}=\Omega/\Omega_{k}$, $\tilde{H}=H/r$, and $\epsilon=kH$. $g=\frac{{\tilde{\chi
}}^{2}}{2\tilde{\Omega}}-2\tilde{\Omega}$, and $\tilde{\chi}=\chi/\Omega_{k}$
where $\chi$ is the epicyclic frequency defined by $ \chi^{2}=2\Omega(2\Omega
+r\frac {\partial \Omega}{\partial r})$. $m$ is the Mach number defined by 
$m=\mid V_{r}\mid/c_{s}$. $q$ is the ratio of advective energy to
viscous dissipated energy, namely
\begin{equation}
C_{v}[\Sigma V_{r}\frac{\partial{T}}{
\partial{r}}-(\Gamma_{3}-1)T(\frac{\partial{\Sigma}}{\partial{t}}+V_{r}\frac
{\partial{\Sigma}}{\partial{r}})]=q \Sigma \nu (r\frac{\partial{\Omega}}{\partial
{r}})^{2}.
\end{equation}
If the disc is radiative cooling dominated, $q$ is nearly zero and if it is 
advection-dominated, $q$ is nearly 1.

By setting the determinants of the coefficients in above perturbed equations 
to zero, we get a dispersion relation:
\begin{equation}
a_{1}{\tilde{\sigma}}^{4}+a_{2}{\tilde{\sigma}}^{3}+a_{3}{\tilde{\sigma}}^{2}+
a_{4}\tilde{\sigma}+a_{5}=0,
\end{equation}
where $a_{i} (i=1,...,5)$ is the coefficients given by
$$
a_{1}=\frac{3}{2},
$$
$$
a_{2}=\alpha[\epsilon^{2}(f+\frac{7}{2})-(\frac{4}{5}+\frac{1}{5}q)g^{2}],$$
$$
a_{3}=\alpha\epsilon g^{2}[\alpha\epsilon(\frac{2}{15}-\frac{7}{15}q)-i
\frac{q}{m}]+\frac{1}{3}(\alpha\epsilon^{2})^{2}(6+7f)+\frac{5}{2}\epsilon^{2}
+\frac{3}{2}{\tilde{\chi}}^{2},$$
$$
a_{4}=2i\alpha^{2}\tilde{\Omega}\epsilon g^{3}\frac{q}{m}+\alpha g^{2}[-
\frac{4}{3}(\frac{4}{5}+\frac{1}{5}q)(\alpha\epsilon^{2})^{2}-i\alpha\epsilon^{3}\frac{q}{m}
+\frac{8}{3}(\alpha\epsilon^{2})^{2}-(\frac{4}{5}+\frac{1}{5}q){\tilde{\chi}}^
{2}$$
$$
+\frac{6}{5}(q-1)
\epsilon^{2}]+
\alpha\epsilon^{2}g[\frac{{\tilde{\chi}}^{2}}{\tilde
{\Omega}}-5\tilde{\Omega}]+\alpha\epsilon^{2}[\frac{4}{3}f(\alpha\epsilon^{2})
^{2}+f{\tilde{\chi}}^{2}+\epsilon^{2}(\frac{5}{2}+f)],$$
$$
a_{5}=(\alpha\epsilon)^{2}[\frac{12}{5}(1-q)\tilde{\Omega}g^{3}+\epsilon^{2}
g^{2}(-\frac{2}{5}+\frac{7}{5}q)+\epsilon^2 g(-\frac{4}{5}-\frac{1}{5}q
-2f\tilde{\Omega})+\epsilon^{4}f].
$$
The stability properties of two inertial-acoustic modes, thermal and viscous 
modes
can be obtained by analyzing the four kinds of solutions of the dispersion 
relation.
The real parts of these solutions correspond to the growth rates of the 
perturbation modes and the imaginary parts correspond to their propagating 
properties. If the growth rate corresponding to certain mode is positive, this 
mode is unstable. Otherwise, it will be stable. Among the four kinds of modes 
in the disc, the stability of thermal and viscous modes have been more 
extensively studied in the literatures. The instability of inertial-acoustic
 mode (or called the pulsational instability), was first addressed in viscous 
 accretion discs by
Kato (1978) and was detailed studied by Blumenthal, Yang \& Lin (1984) later. It 
has been suggested that the inertial-acoustic instabilities may account for the 
observed quasi-period oscillations (QPO) in some Galactic black hole candidates
 (Chen \& Taam 1995).

\section{Numerical results}

In this section we will numerically solve the dispersion relation Eq. (13). 
The rotation law of disc is further assumed to be Keplerian. Although Narayan
\& Yi (1994) obtained a sub-Keplerian rotation law ($\Omega \sim 0.5 \Omega_k$
in a hot optically thin disc), such a difference will not affect significantly
on our results.
According to the different possible structure of a two-temperature disc, we 
solve the dispersion relation in following three cases:\\
(a) {\it Geometrically thin, cooling dominated two-temperature disc}.  
We take $\tilde{\Omega}=\tilde{\chi}=1$,
$\alpha=0.01, m=0.01$. According to the local restrictions, $\lambda/H$ is set 
from 1 to 80 for a geometrically thin disc (in this section we take $H/r=0.01$). 
By solving the dispersion 
relation, we get the results shown in Fig.1. For comparison, we also show the 
stability properties of an optically thin, geometrically thin bremsstrahlung 
disc, which has been suggested to be thermally unstable (Pringle, Rees \& 
Pacholczyk 1973). Fig. 1(a) shows the case without advection. We can clearly 
see that the thermal mode is always unstable and the Lightman \& Eardley 
viscous mode is always stable. This result agree well with the previous 
finding. The inertial-acoustic modes are slightly unstable but stable to very 
short wavelength perturbations.  In addition, we see that the thermal mode in a
 bremsstrahlung disc is 
more unstable than that in a two-temperature disc. Fig. 1(b)
 shows the case with very little advection ($q=0.01$). It is clear that the 
 inclusion of very little advection has nearly no effects on the thermal and 
 viscous modes, but leads to the departure of two acoustic modes. Comparing
 to the case without advection, the 
 outward propagating acoustic mode (hereafter O-mode) now becomes more unstable, 
 whereas the inward propagating acoustic mode (hereafter I-mode) becomes more 
 stable. In Fig. 1(c) we show the case with $q=0.01$ and $m=0.1$. Comparing 
 with Fig. 1(b), the increase of Mach number leads to the decrease of the 
 departure of two acoustic modes. Actually, as we have noted in a previous 
 work (Wu \& Li 1996), such a departure is proportional to the term $q/m$. 
 Both the O-mode and the I-mode are unstable to the longer wavelength perturbation 
  if $q/m$ less than 0.1. The change of $q/m$, however, has nearly 
 no effects on the thermal and viscous modes. Fig. 1(d) shows the case when  
 $q=0.01$, $m=0.01$ but with thermal diffusion included. Comparing with Fig 
 1.(b), we see that the inclusion of thermal diffusion has nearly no effect 
 on the viscous mode and two acoustic modes, but it tends to stabilize the 
 thermal instability in the short perturbation wavelength case. In general, 
 if the disc is geometrically thin and cooling dominated, the stability 
 properties of a two-temperature disc are quite similar to those of a 
 bremsstrahlung disc. But we still see clearly that the thermal mode in a 
 bremsstrahlung disc is more unstable than that in a two-temperature disc. 
 We noted that these results agree well with those obtained by Luo \& Liang 
 (1994), who have pointed out that the thermal mode is always unstable and 
 viscous modes is always stable in a hot optically thin disc. However, they 
 did not consider the stability of inertial-acoustic modes and the effects 
 of radial viscous force, advection and thermal diffusion.\\
(b) {\it Geometrically slim, cooling dominated two-temperature disc}. Some 
previous works have suggested that a hot optically thin disc may be not 
geometrically thin but geometrically slim or thick (e.g., SLE) even if the 
advection term was ignored. Here we show the stability of a geometrically slim,
 two-temperature disc. In this section we take $H/r=0.6$. The perturbation 
 wavelength is set from 0.2 to 2 according to the local approximation. 
 Fig. 2(a) shows the case when $m=0.01$, $\alpha=0.001$ and without advection 
 ($q=0$). We can see that the inertial-acoustic modes and the viscous mode are 
 always stable. Only the thermal mode is unstable when perturbation wavelength 
 is longer than 1.5H. Fig. 2(b) show the similar results as Fig. 2(a) but with very 
 little advection included ($q=0.01$). In comparison with Fig. 2(a) we see 
 that two acoustic modes now slightly depart from each other. For clearity, 
 Fig. 2(c) compares the stability of thermal modes in two cases above. The 
 stability of thermal mode of an optically thin, bremsstrahlung disc is also 
 shown for comparison. It is quite clear that the inclusion of very little 
 advection has only very slight effects on the thermal mode, which is always 
 unstable in a bremsstrahlung disc but can become stable in a two-temperature 
 disc if the perturbation wavelength is shorter. The case with thermal 
 diffusion considered is shown in Fig. 2(d). We clearly see that all four 
 kinds of modes in a two-temperature disc are always stable if the thermal 
 diffusion is considered. We also noted that  an optically thin, 
 geometrically slim bremsstrahlung disc can become stable if the thermal 
 diffusion is considered, even if it is more thermally unstable than a 
 two-temperature disc when the thermal diffusion is ignored.\\
(c) {\it Geometrically slim, advection-dominated two temperature disc}. 
The advection-dominated equilibrium of an optically thin, two-temperature disc 
has been recently constructed by some authors (e.g., Narayan \& Yi 1995b; 
Chen et al. 1995; Nakamura et al. 1996). Such a disc is usually geometrically slim. In Fig. 3, we show its 
stability against short wavelength perturbations. Here we take $q=0.99$ and $H/r=0.6$.
 Fig. 3(a) shows
 the case with $m=0.01$ and $\alpha=0.001$. In this case, the thermal, viscous 
 modes are always stable. The inertial-acoustic modes depart from each other. 
 The I-mode is stable but the O-mode can become unstable if the perturbation 
 wavelength is larger. Fig. 3(b) shows the case with different Mach number 
 ($m=0.1$). Comparing with Fig. 3(a), the departure of two acoustic modes 
 becomes less due to the decrease of $q/m$. The O-mode can become stable if
  $q/m$ less than 10. However, the change of $q/m$ has no effects on the 
  thermal and viscous modes, which are always stable. In Fig. 3(c), we show 
  the case similar to Fig. 3(a) but with thermal diffusion included. We can 
  clearly see that the inclusion of thermal diffusion has a significant effect 
  to stabilize the thermal mode, but it has nearly no effects on the viscous 
  and acoustic modes. We noted that in an optically thin, advection-dominated 
  disc, the stability properties do not change evidently if the different radiative 
  cooling mechanism was involved. Whether the disc is bremsstrahlung one or 
  two-temperature one, the stability properties are always the same. This is, 
  of course, due to the unimportance of radiative cooling in an 
  advection-dominated disc.

\section{Discussions}

The stability of a hot optically thin two-temperature disc has been discussed 
in above sections by numerically solving the dispersion relation. We proved 
that the thermal instability exists in the disc if it is cooling dominated but 
dispears if it is advection-dominated. The Lightman-Eardley viscous instability
 is always absent in a hot optically thin disc. These properties agree 
 well with previous qualitative results. We have also investigated 
 the stability of two inertial-acoustic modes, which has not been discussed 
 previously in a hot optically thin, two-temperature disc. In a geometrically 
 thin, radiative cooling dominated disc, we found that the O-mode is always 
 unstable but the I-mode is unstable only if the term $q/m$ less than 0.1. If 
 the disc is cooling dominated but geometrically slim, no inertial-acoustic 
 mode is unstable. However, in a geometrically slim and advection-dominated  
 disc, the O-mode can become unstable if $q/m$ larger than 10, though the 
 I-mode is always stable. These stability properties of acoustic modes may
 be important when they are involved to explain the QPO phenomena in some
 systems such as Galactic black hole candidates (Chen \& Taam 1995;
Manmoto
et al. 1996).

A simple comparison shows that the bremsstrahlung disc is more thermally 
unstable than the two-temperature disc if they are cooling dominated. This 
can been seen clearly by comparing their different temperature-dependence of 
the cooling rates. In a bremsstrahlung disc, the cooling rate ${Q_b}^-$ is 
proportional to ${\Sigma}^2$. In a two-temperature disc, the cooling rate 
${Q_t}^-$ is proportional to $\Sigma^{7/5}T^{1/5}$. In these two cases the 
energy generated rate by viscous dissipation can be both expressed as 
$Q^+=\Sigma \nu \Omega \frac{\partial \Omega}{\partial r} \propto \Sigma T$. 
Thus, if we give a positive perturbation $\delta T$ and keep the $\Sigma$ 
unperturbed, $Q^+$ will grow more rapidly than ${Q_b}^-$ and ${Q_t}^-$, which 
leads to thermal instability (see also Piran 1978). Because $Q^+-{Q_t}^-$ is
less than 
 $Q^+-{Q_b}^-$ with a positive perturbation of temperature, the 
two-temperature disc is less thermally unstable than the bremsstrahlung disc.
 If the hot optically thin disc is advection-dominated, however, the disc 
 stability will not depend on the detailed cooling mechanism due to its 
 less importance.

It is quite evident that the advection and thermal diffusion have significant 
effects on the stability of a hot optically thin disc. These effects are more 
evident if the disc is not geometrically thin. As having been pointed out by 
Wu \& Li (1996), the thermal diffusion in an accretion disc is in proportional
 to $(H/r)^{2}$. Therefore, the thermal diffusion should be seriously 
 considered in the hot optically thin disc which is usually geometrically slim 
 or thick. We think that the inclusion of thermal diffusion will affect not 
 only the stability but also the structure of accretion discs. A future 
 investigation on this point is expected.

More recently, we noted that Kato et al. (1997) pointed out that the perturbations could be also local in the vertical direction for a geometrically thick disc. This is true but the consideration of the vertical local perturbations seems to have only minor contribution to the growth rate of unstable mode. For a two
temperature disc, the independent analytic stability analysis of Yamasaki (1997) found that there are two kinds of modes in the disc. One is thermal mode which is slightly unstable and the other is viscous Lightman \& Eardley mode which is always stable. The inclusion of thermal diffusion will stabilize the thermal mode. These result agree quite well as our quantitative analyses. However, we noted that the radial viscous force was not included in the analysis of Yamasaki (1997). Consireation of this will contribute a term in proportional to $(kH)^2$ in the perturbed equation of radial momentum conservation, which can not be neglected for a geometrically thick disc. It will result in the absence of the weak thermal instability of a hot optically thin disc when the short wavelength perturbations are considered (Wu, Yang \& Yang 1994; Wu \& Li 1996). 

We should noted that the radiative cooling rate of a two-temperature 
disc may be not as simple as we assumed in Section 2. For example, we have 
assumed that the radiation energy density of soft photons, $U_r$, is unchanged 
on dynamical timescale, which may be not always the case. Some recent works on 
the hot optically thin discs suggested that the cooling mechanism is rather 
complicated (See e.g., Esin et al. 1996). The disc structure also depends on 
the radius significantly. This implies that the disc stability may be quite 
different from one radius to another. Thus, the global stability analyses of 
a hot optically thin disc are still needed. In addition, the pair production 
and
annihilation in
a hot accretion disc are totally neglected in this paper for simplicity because
 the inclusion of pair process will make 
the linear stability analysis very complicated.  
Although this process may be important especially if the disc 
temperature is higher enough, the pair density is believed to be not too high
and their influence is limited (Bjornsson et al. 1996; Kusunose \& Mineshige 1996). However, we still expect a 
future detailed work 
would 
be done to investigate their influences on the structure and stability of a hot
accretion disc.

Together with some previous detailed stability analyses, our study show that
there are perhaps only two stable thermal equilibria of accretion discs. One 
is optically
thin, advection-dominated and the other is optically thick, gas-pressure
and radiative cooling dominated. These equilibria are probably related with some stable, inactive 
astronomical systems. For example, the optically thin, advection-dominated
accretion discs have been suggested to exist at the center of our
Galaxy (Narayan, Yi \& Mahadevan 1995), nearby elliptical 
galaxies 
(Fabian \& Rees 1995), fainter 
AGNs (Lasota et al. 1996) and soft X-ray transient sources in the quiescent
state (Narayan, McClintock \& Yi 1996). On the other hand, the unstable thermal
equilibria of accretion discs, such as the optically thin, radiative cooling
dominated one, the optically thick, radiation pressure dominated one and the
optically thick, advection-dominated one, probably exist in some unstable
systems such as the inner region of AGNs, X-ray binaries and cataclysmic 
variables. In order to understand the light variability of these systems more 
clearly, the detailed time-dependent nonlinear studies on the evolution of their disc 
structures are still expected.

\section*{Acknowledgments}
I am very grateful to Professor Ramesh Narayan for mentioning me a few
important
references and some valuable suggestions.
I also thank Professor Qibin Li and Dr. Yongheng Zhao for many helpful discussions. 
The work was partially supported by the Postdoc Science Foundation of China.\\


\section*{References}

Abramowicz, M.A., Czerny, B., Lasota, J.P., Szuszkiewicz, E. 
1988, ApJ, 332, 646\\
 Abramowicz, M.A., Chen, X., Kato, S., Lasota, J.-P., Ragev, O. 
1995, ApJ, 438, L37\\
Bjornsson, G., Abramowicz, M., Chen, X., Lasota, J.-P., 1996, ApJ,
467, 99\\
 Blumenthal, G.R., Yang, L.T., Lin, D.N.C. 1984, ApJ, 287, 774\\
 Chen, X. 1995, MNRAS, 275, 641\\
 Chen, X. 1996, in ``Basic Physics  of Accretion Disks" ed. by 
Kato S. et al.,
Gondon and Breach Science Publishers, in press\\
 Chen, X., Abramowicz, M.A., Lasota, J.-P., Narayan, R., Yi, I. 
1995, ApJ, 443, L61\\
 Chen, X., Taam, R.E. 1995, ApJ, 441, 354\\
 Esin, A.A., Narayan, R., Ostriker, E., Yi, I.,1996, APJ, 465, 312\\
 Fabian, A.C., Rees, M.J., 1995, MNRAS, 277, L55\\
 Kato, S. 1978, MNRAS, 185, 629\\
 Kato, S., Abramowicz, M.A., Chen, X. 1996, PASJ, 48, 67\\
 Kato, S., Honma, F., Matsumoto, R. 1988, MNRAS, 231, 37\\
Kato, S., Yamasaki, T., Abramowicz, M.A., Chen, X., 1997, PASJ, 49, 221\\
 Kusunose, M. Mineshige, S., 1996, ApJ, 468, 330\\
 Kusunose, M., Takahara, F., 1988, PASJ, 40, 709\\
 Kusunose, M., Takahara, F., 1990, PASJ, 1, 263\\
 Lasota, J.P., Abramowicz, M., Chen, X., Krolik, J., Narayan, R., 
Yi, I., 1996, ApJ, 462, 142\\
 Lightman, A., Eardley, D. 1974, ApJ, 187, L1\\
 Luo, C., Liang, E.P., 1994, MNRAS, 266, 386\\
Manmoto, T., Takeuchi, M., Mineshige, S., Matsumoto, R., Negoro, H.,
1996, ApJ, 464, L135\\
 Misra, R., Melia, Fulvio., 1996, ApJ, 465, 869\\
 Nakamura, K.E., Matsumoto, R., Kusunose, M., Kato, S., 1996, PASJ, 48, 761\\
 Narayan, R., McClintock, J.E., Yi, I., 1996, ApJ, 457, 821\\
 Narayan, R., Yi, I. 1994, ApJ, 428, L13\\
 Narayan, R., Yi, I. 1995a, ApJ, 444, 231\\
 Narayan, R., Yi, I. 1995b, ApJ, 452, 710\\
Narayan, R., Yi, I., Mahadevan, R., 1995, Nature, 374, 623\\
 Paczy\'nski, B., Wiita, P.J., 1980, A\&A, 88, 23\\
 Papaloizou, J.C.B., Stanley, G.Q.G. 1986, MNRAS, 220, 593\\
 Piran, T. 1978, ApJ, 221, 652\\
 Pringle J.E., 1976, MNRAS, 177, 65\\
 Pringle, J.E., Rees, M.J., Pacholczyk, A.G. 1973, A\&A, 29, 179\\
 Shakura, N.I., Sunyaev, R.A. 1973, A\&A, 24, 337\\
 Shakura, N.I., Sunyaev, R.A. 1976, MNRAS, 175, 613\\
 Shapiro, S.L., Lightman, A.P., Eardley, D.N. 1976, ApJ, 204, 187\\
 Spitzer, L., 1962, {\it Physics of fully ionized gases}, Wiley, 
New York.\\
 Wandel, A., Liang, A.P., 1991, ApJ, 380, 84\\
 Write, T.R., Lightman, A.P., 1989, APJ, 340, 1024\\
  Wu, X.B., Li, Q.B., ApJ, 469, 776\\
Wu, X.B., Yang, L.T., Yang, P.B., 1994, MNRAS, 270, 465\\
Yamasaki, T., 1997, PASJ, 49, 227

\newpage
\begin{center}
{\bf FIGURE CAPTIONS}\\
\end{center}
\noindent
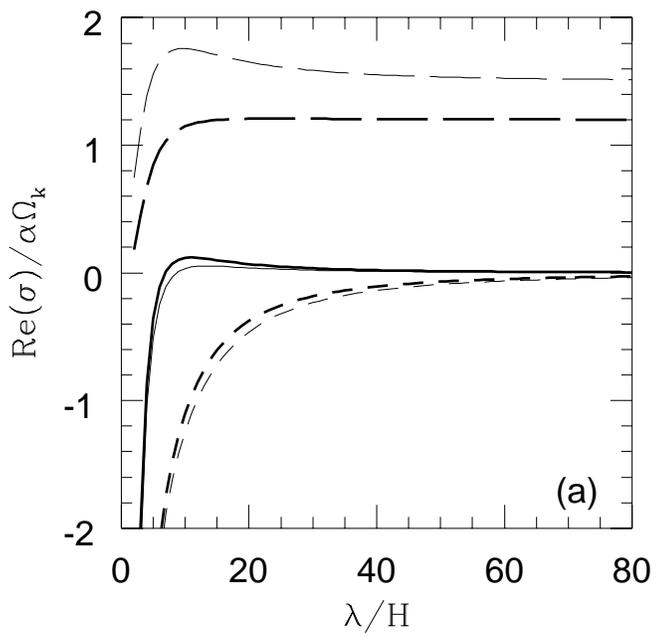
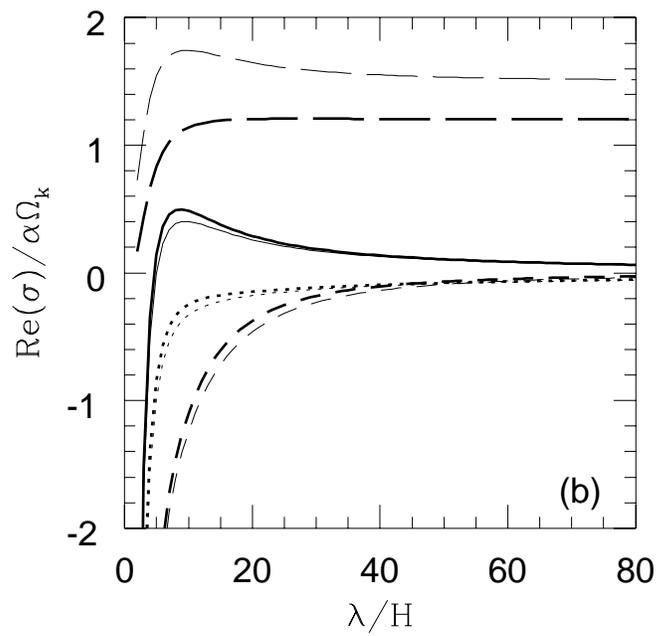
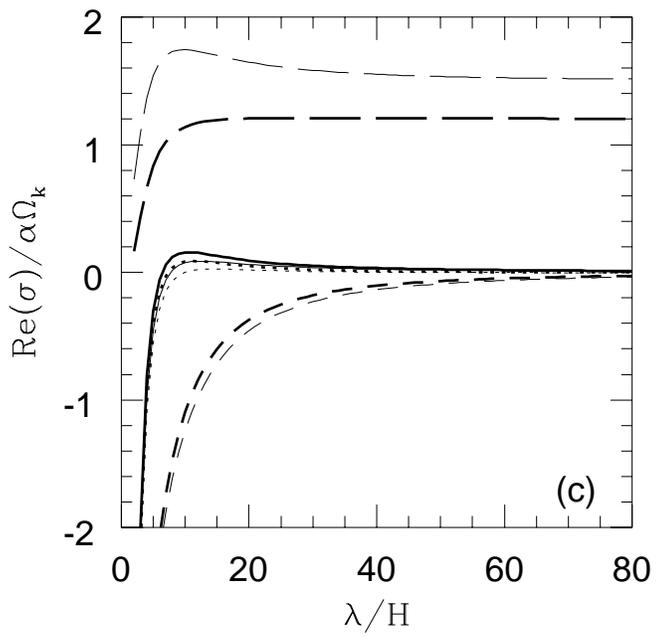
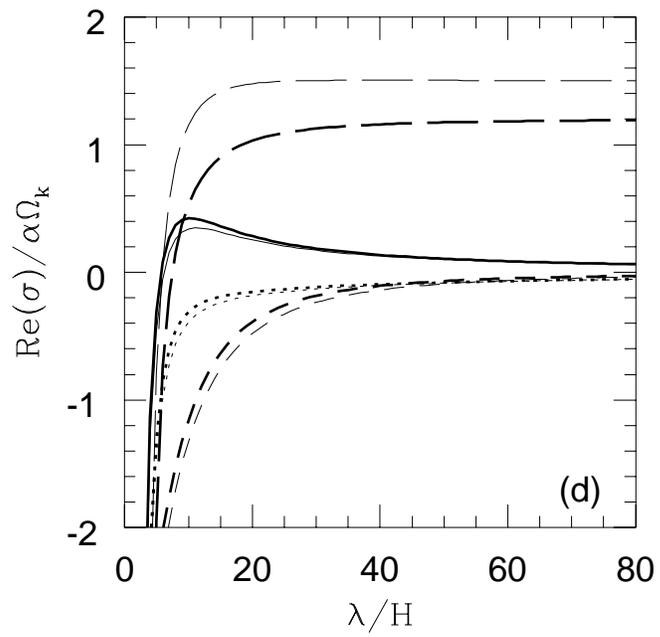
 {\bf Figure 1.} Stability of a geometrically thin, optically thin disc. Heavy 
 lines 
 represent the case of a two-temperature disc and light lines represent the case
  of bremsstrahlung disc. In (a) no advection is included, the solid, long-dash
   and short-dash lines correspond to the acoustic modes, thermal mode and 
   viscous mode respectively. (b), (c) and (d) show the cases with very little 
   advection (q=0.01), parameters (m, f) are taken as (0.01, 0) in (b), 
   (0.1, 0) in (c) and (0.01, 3) in (d). The long-dashed and short-dashed 
   lines in (b), (c) and (d) have the same meaning as in (a) but the solid 
   and dotted lines 
   correspond to the O-mode and the I-mode.\\[3mm]

\noindent
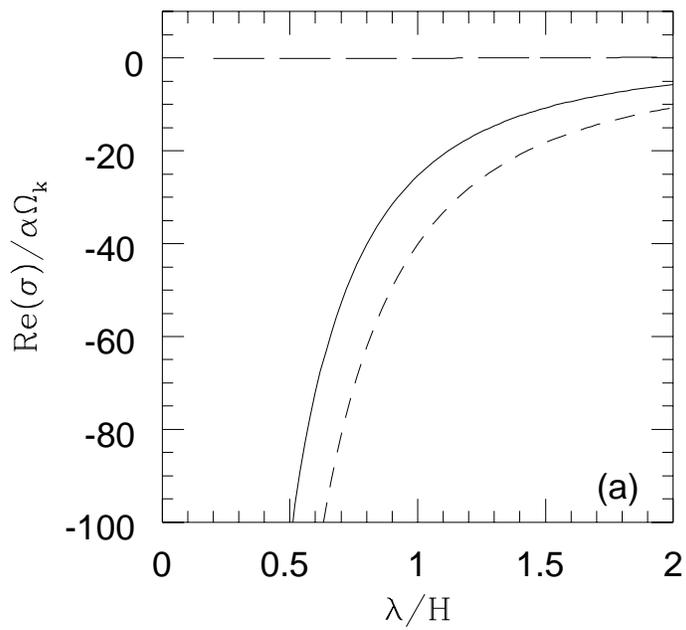
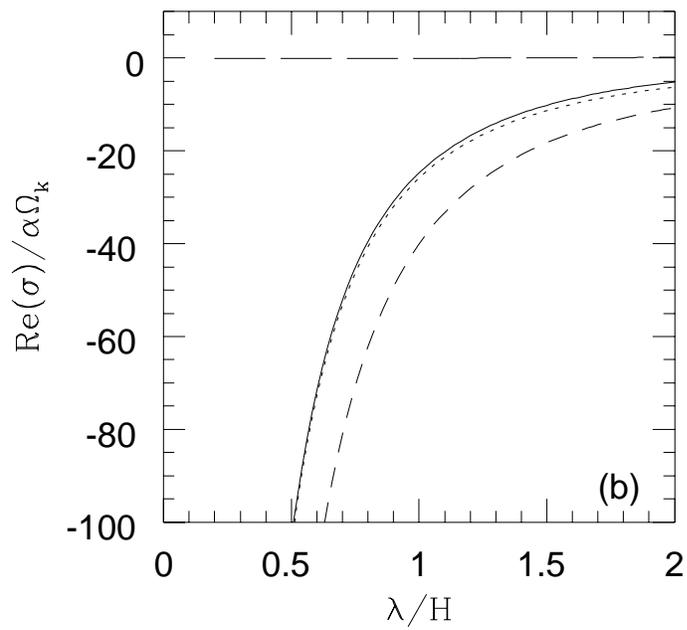
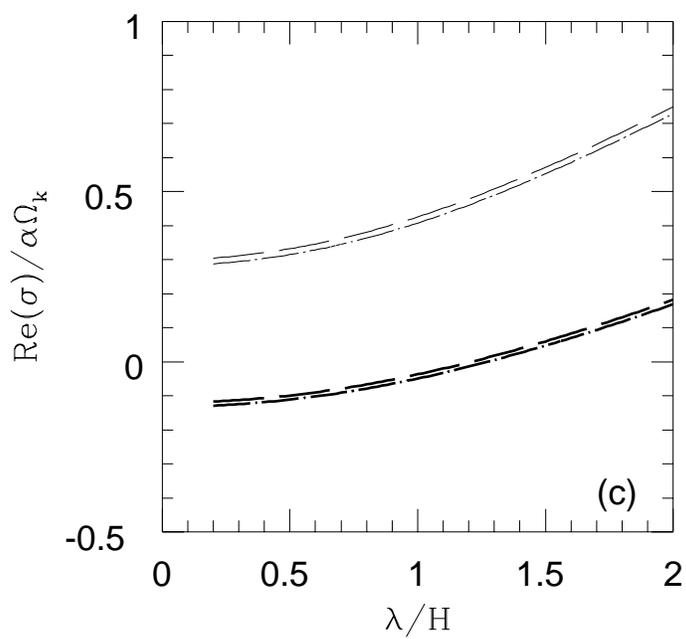
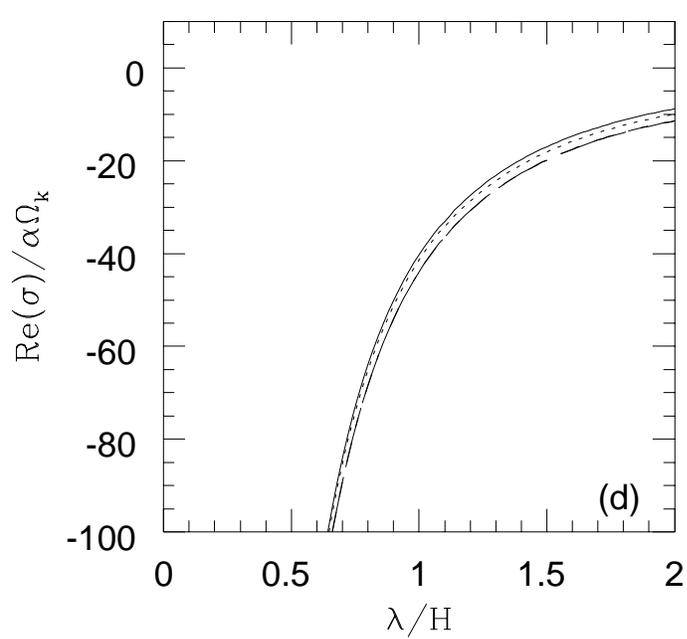
{\bf Figure 2.} Stability of a geometrically slim, hot two temperature disc. In 
(a) no advection is included, the solid, long-dashed and short-dashed lines 
correspond to the acoustic modes, thermal mode and viscous mode respectively. 
(b) shows the case with very little advection but without thermal diffusion 
(q=0.01, f=0). The  solid and dotted lines correspond to the O-mode and the 
I-mode. In (c) the differences of thermal instabilities in two kinds of 
optically thin discs are indicated. The heavy lines represent the case of 
a two-temperature disc and light lines represent the case of bremsstrahlung 
disc. The dot long-dashed lines correspond to the thermal modes when little 
advection is included (q=0.01). (d) shows the case similar with (b) but with thermal 
diffusion included (f=3). The thermal and viscous modes are mixed together in 
this case.\\[3mm]

\noindent
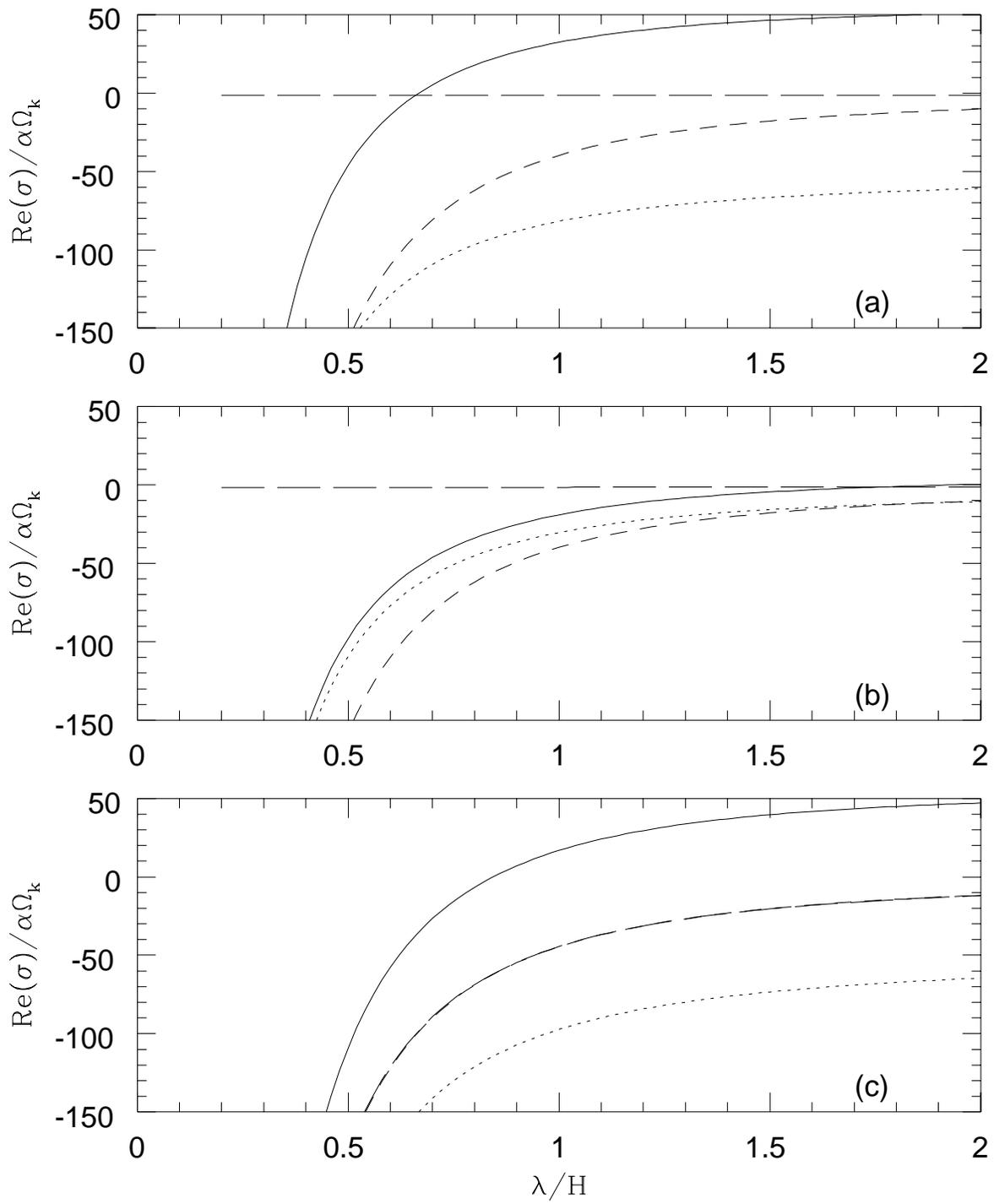
{\bf Figure 3.} Stability of a geometrically slim, advection-dominated two 
temperature disc. The solid, dotted, long-dashed and short-dashed lines 
correspond to the O-mode, I-mode, thermal and viscous mode respectively. 
The parameters (m, f) are (0.01, 0) in (a), (0.1, 0) in (b) and (0.01, 3) 
in (c). The thermal and viscous modes are mixed together in (c) where the thermal 
diffusion is included.

\end{document}